\documentclass{article}
\usepackage{spconf,amsmath,graphicx}

\usepackage{enumitem}
\usepackage{graphicx}
\usepackage{subcaption}
\setlist{nosep, leftmargin=14pt}
\usepackage{etoolbox}
\usepackage{multirow}
\usepackage{xcolor}
\usepackage{tikz}
\usepackage{url}
\apptocmd{\thebibliography}{\setlength{\itemsep}{0pt}\setlength{\parskip}{0pt}}{}{}
\usepackage{mwe} 

\title{Biomechanically Accurate Gait Analysis: A 3D Human Reconstruction Framework for Markerless Estimation of Gait Parameters}
\name{Author(s) Name(s)\thanks{Some author footnote.}}
\address{Author Affiliation(s)}
\name{Akila Pemasiri$^{\star }$,  Ethan Goan$^{\star}$,  Glen Lichtwark$^{\star}$, Robert Schuster$^{\dagger}$,  Luke Kelly$^{\dagger}$,  Clinton Fookes $^{\star}$  }

\address{$^{\star}$ Queensland University of Technology, Brisbane
    $^{\dagger}$ Griffith University, Gold Coast  }
\begin{document}
%
\maketitle
\begin{abstract}
This paper  presents a biomechanically interpretable framework for gait analysis using 3D human reconstruction from video data. Unlike conventional keypoint based approaches, the proposed method extracts biomechanically meaningful markers analogous to motion capture systems and integrates them within OpenSim for joint kinematic estimation. To evaluate performance, both spatiotemporal and kinematic gait parameters were analysed against reference marker-based data. Results indicate strong agreement with marker-based measurements, with considerable improvements when compared with pose-estimation methods alone. The proposed framework offers a scalable, markerless, and interpretable approach for accurate gait assessment, supporting broader clinical and real world deployment of vision based biomechanics.\footnote{This paper has been accepted for ISBI 2026.}
\vspace{-5pt}
\end{abstract}
\begin{keywords}
Gait analysis, Biomechanical modeling, 3D human pose reconstruction \vspace{-5pt}
\end{keywords}
\section{Introduction}
\label{sec:intro}
\vspace{-10pt}
Gait analysis plays a vital role in clinical diagnostics, rehabilitation monitoring , and biomechanical research \cite{hii2023automated, shin2021quantitative,marin2020my}. It provides quantitative insights into human movement, assisting clinicians and researchers in assessing neuromuscular disorders, balance impairments, and functional recovery \cite{hii2023automated, shin2021quantitative,marin2020my}. 
Traditionally, gait analysis relied heavily on visual observation, where trained clinicians interpreted movement patterns qualitatively \cite{lord1998visual}. However such subjective approaches lack precision and reproducibility. To overcome these limitations, marker-based motion capture systems became the gold standard, enabling high-fidelity measurement of three-dimensional motion through reflective markers tracked by infrared cameras \cite{raghu2021kinematic,goldfarb2021open}. Complementary technologies such as force plates \cite{zeni2008two}, pressure sensors \cite{park2018development}, and inertial measurement units (IMUs) \cite{mobbs2022gait} further enhanced data richness, allowing spatiotemporal and kinematic analysis.

However, these systems present several practical drawbacks: they are expensive, require controlled laboratory environments, and depend on precise marker placement by trained personnel \cite{eguchi2025gait,d2020markerless,panconi2025deeplabcut}. Moreover, the presence of extra instruments such as markers,  force platforms can alter natural movement patterns. For clinical  gait assessment, especially among older adults or individuals with mobility impairments such constraints limit scalability and routine use.

\begin{figure}[t]
    \centering
    \begin{subfigure}[b]{0.48\columnwidth}
        \centering
        \includegraphics[height=3.0cm]{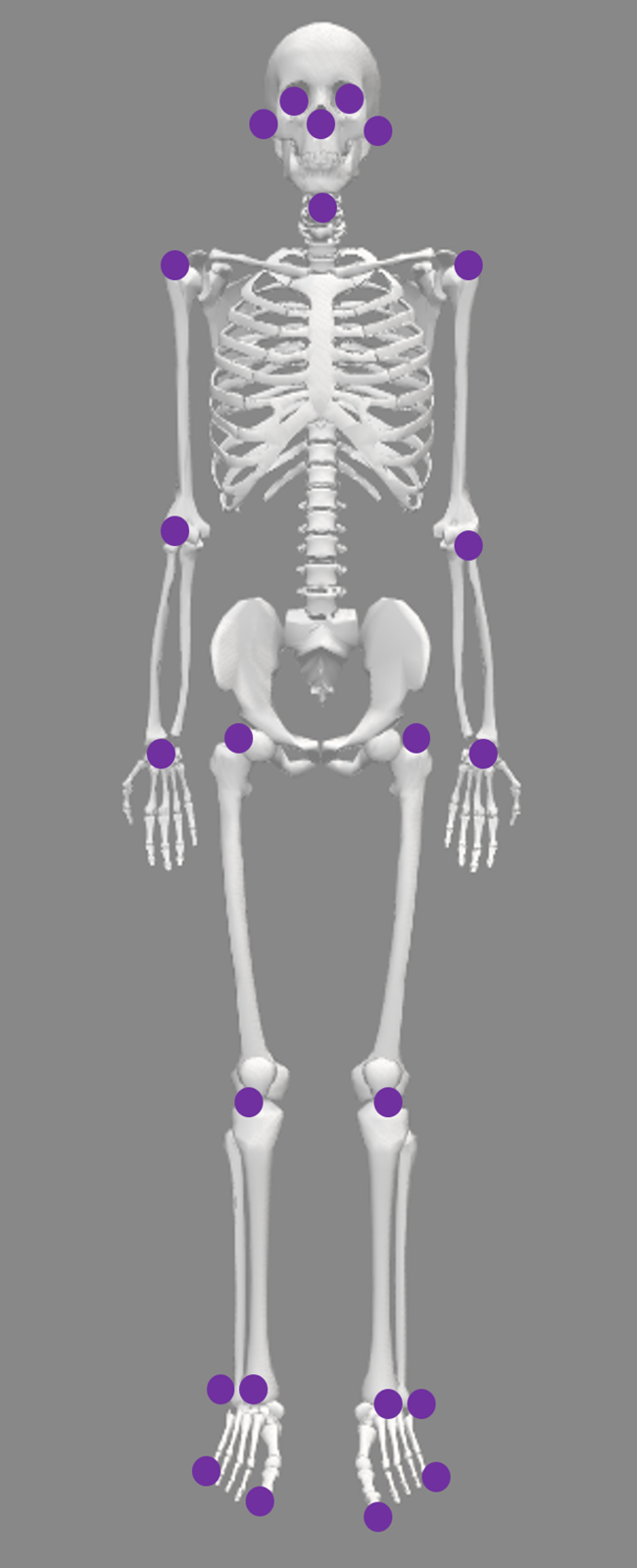}
        \caption{Pose estimation  model}
        \label{fig:common_landmarks}
    \end{subfigure}
    \hspace{0.02\columnwidth} 
    \begin{subfigure}[b]{0.48\columnwidth}
        \centering
        \includegraphics[height=3.0cm]{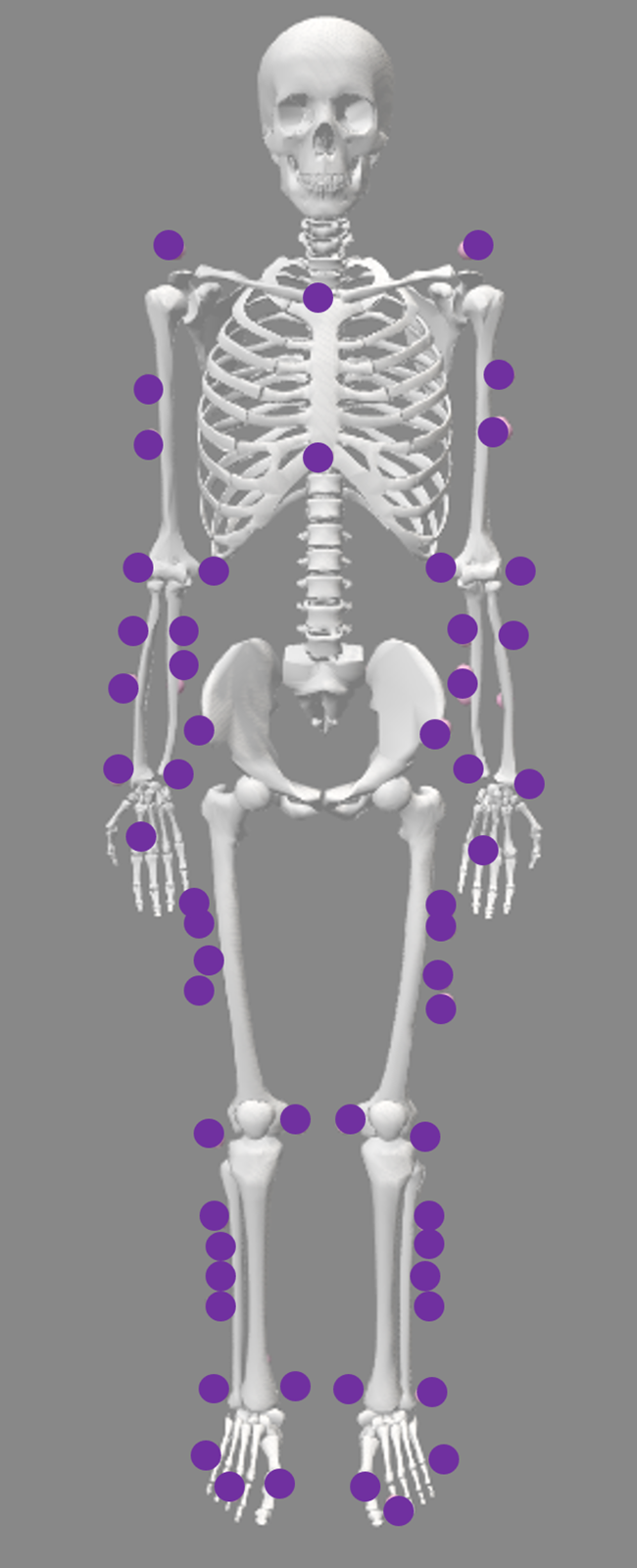}
        \caption{Biomechanical model \cite{rajagopal2016full}}
        \label{fig:opensim_landmarks}
    \end{subfigure}
    \vspace{-18pt}
    \caption{Comparison of body landmarks obtained from the pose estimation \cite{cao2019openpose} and biomechanical model \cite{rajagopal2016full}. \vspace{-20pt} }
    \label{fig:common_vs_anatomical}
\end{figure}

Recent advances in vision-based pose estimation have sparked growing interest in markerless gait analysis \cite{eguchi2025gait,d2020markerless,panconi2025deeplabcut}. These methods leverage deep learning to infer body keypoints from video, offering a cost effective and non intrusive alternative to traditional systems. However, most existing pose estimation models (e.g., COCO \cite{lin2014microsoft}) employ a predefined set of anatomical keypoints optimized for generic computer vision tasks, not biomechanical accuracy (Fig \ref{fig:common_vs_anatomical}). Consequently, the extracted keypoints may not align with clinically relevant joint centers or musculoskeletal landmarks, introducing errors when used for gait parameter computation.

We propose a framework for biomechanically accurate gait analysis based on 3D human reconstruction. Instead of relying on sparse and generic 2D keypoints, our approach reconstructs a detailed 3D representation of the human body and extracts biomechanically relevant markers analogous to those used in motion capture systems. These markers can then be integrated with biomechanical modeling tools such as OpenSim \cite{delp2007opensim} to estimate gait parameters with improved accuracy and interpretability.
\begin{figure*}[!ht]
    \centering
    \includegraphics[width=0.95\linewidth]{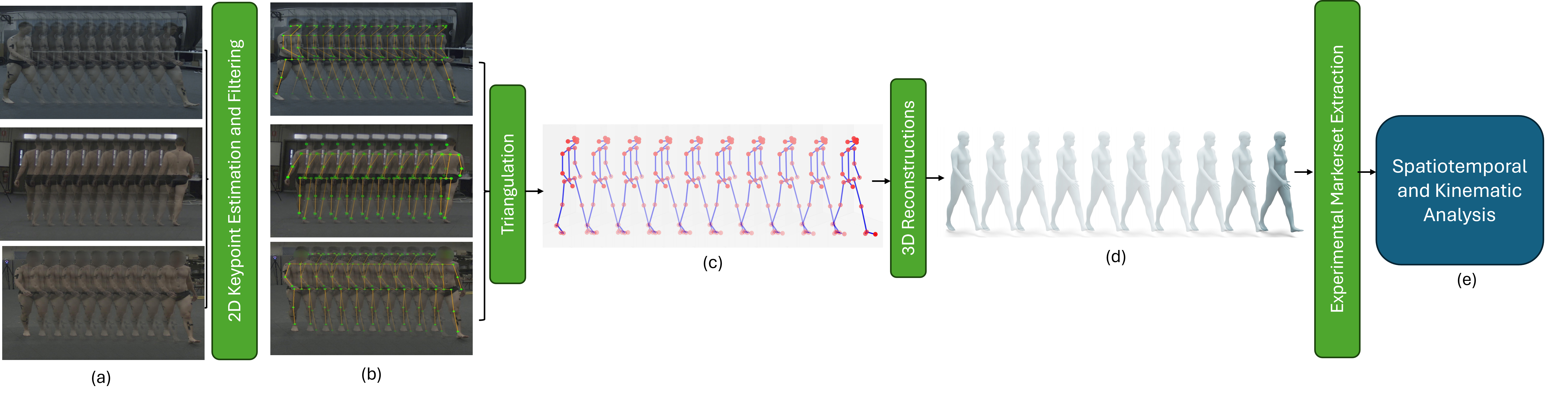}
    \caption{Overview of the gait analysis framework. The pipeline consists of five main stages: 2D pose estimation, 3D triangulation, body shape estimation and 3D reconstruction, extraction of anatomical markers, and gait parameter estimation. Gait parameters are derived both manually and through musculoskeletal modeling approaches such as OpenSim \cite{delp2007opensim}.}
    \label{overview}
\end{figure*} 
It should be noted that this work focuses on an interpretable biomechanical modeling pipeline, rather than purely data driven gait parameter estimation, to ensure transparency and scientific validity in gait analysis.
The proposed framework offers three major advantages:

\begin{enumerate}
    \item Biomechanically accurate markers leading to more realistic and precise estimation of gait features.
    \item Integration with musculoskeletal models enables deeper understanding of joint kinematics and dynamics beyond surface-level motion.
    \item  The approach overcomes limitations of marker based systems, providing a scalable and non-intrusive,  pathway for gait assessment in clinical and real world environments.     \vspace{-10pt}
\end{enumerate}

\section{Method}
 \vspace{-10pt}
\subsection{3D Shape Estimation}
 \vspace{-5pt}
The overall framework of the proposed methodology is depicted in Fig. \ref{overview}. Synchronized views from multiple cameras were used to capture motion of the subject. Using the well established and commonly used pose estimation methods, keypoints corresponding to the widely used body landmarks were obtained (Fig. \ref{fig:common_vs_anatomical}a).The subject of interest was tracked throughout the sequence using a robust multi-object tracking framework \cite{varghese2024yolov8}.

Once the 2D keypoints were filtered, they were triangulated across multiple camera views to reconstruct 3D coordinates for each landmark. This allowed for precise estimation of the subject’s spatial configuration while accounting for potential occlusions and noise in individual camera views. Subsequently, the reconstructed 3D keypoints were processed using the EasyMocap \cite{easymocap,dong2021fast} framework to estimate the parametric 3D body shape and pose parameters relevant to the SMPL \cite{SMPL:2015} model. 

Finally, a biologically accurate marker set (referred  as ``Experimental markers'' within) was extracted from the estimated 3D body shape, representing anatomical landmarks aligned with musculoskeletal modeling conventions. This marker set enables subsequent biomechanical analyses, such as motion tracking,  and musculoskeletal simulations, while ensuring anatomical fidelity. \vspace{-10pt}

\subsection{Gait Parameter Estimation}
\vspace{-5pt}
In this study, we compute both spatiotemporal gait parameters and kinematic parameters from the captured motion data. Unlike most existing studies that focus on either spatiotemporal or kinematic analysis in isolation, our approach integrates both types of parameters to provide a more comprehensive characterization of gait behaviours.

\noindent \textbf{Spatiotemporal parameters:} The spatiotemporal parameters analysed include stride length, stride time, step length and step time. For real marker data and the markers obtained by our proposed method, these parameters were computed using the markers associated with the right and left heels, while for pose-only method, keypoints 21 (left heel) and 25 (right heel) were used. 

\noindent \textbf{Kinematic parameters:}  Experimental markers from the static pose were used to perform the scaling of the musculoskeletal model in OpenSim. 
The Scaling Tool in OpenSim was used to adjust the generic model dimensions to align virtual and experimental markers for each participant. Body mass and stature were incorporated into the scaling process to ensure accurate model personalization. Using the generated experimental markers, inverse kinematics was then performed on the scaled model to estimate the joint angles throughout the motion.

\section{Experiments and Results}
\vspace{-10pt}
\subsection{Dataset}
\vspace{-5pt}
We used the BioCV dataset \cite{evans2024biocv}, which contains synchronized multi-view video recordings and 3D pose annotations of human subjects performing various gait and motion activities. The dataset provides calibrated camera parameters and groundtruth 3D joint locations, enabling accurate evaluation of pose estimation and the proposed gait analysis. 

\subsection{2D Pose Estimation Methods}
\vspace{-5pt}
To extract 2D human joint locations from the multiview video frames, we employed three well established pose estimation frameworks: OpenPose \cite{cao2019openpose}, MMPose \cite{mmpose2020}, and a YOLO based pose estimation model \cite{varghese2024yolov8}.  We selected these pose estimation methods for this study, as these frameworks have been widely adopted in previous research on gait and human motion analysis \cite{hii2023automated, shin2021quantitative,marin2020my}.

Each method was applied independently to all camera views to generate 2D keypoints for the target subject. The resulting 2D keypoints were refined through temporal filtering before being used for 3D reconstruction and subsequent gait parameter analysis. This enabled a comparative evaluation of pose estimation performance across different methods and their influence on derived gait metrics.
\vspace{-5pt}

\vspace{-5pt}
\subsection{Gait Parameters}
\vspace{-5pt}
In this section we provide the  dual level analysis of spatiotemporal and kinematic gait parameters, contrasts with the majority of prior works, which typically examine only one class of parameters. By jointly analyzing spatiotemporal, and joint kinematic information, our framework enables a more holistic assessment of gait patterns.

\noindent \textbf{Spatiotemporal gait parameters:}  Spatiotemporal gait parameters were computed using three different sources of markers: real markers, synthetic markers generated by our method, and heel markers obtained from the pose estimation process. To evaluate the alignment of the gait parameters derived from the different methods, the correlation between the spatiotemporal values obtained from the real markers and those obtained from the other two methods was calculated. The mean absolute error (MAE) was used to quantify the magnitude of deviation from the reference measurements. A high correlation indicates strong agreement between our method and the real marker-based measurements, while a low MAE reflects improved accuracy and consistency in estimating the gait parameters.

Results of this analysis are summarized in Table \ref{table:spatiotemporal_results1}, which demonstrates the proposed method consistently outperforming the 3D Pose-only approach across all parameters, achieving higher correlation values and lower mean absolute errors (MAE) with respect to reference marker-based measurements. Among the 2D estimators, OpenPose provided the most accurate results, with the proposed method achieving correlation values up to 0.98 and MAE as low as 0.0117 for step length estimation. YOLO and MMPose also showed notable improvements under the proposed framework, though their performance remained slightly below that of OpenPose. Overall, the results indicate that the proposed method enhances the robustness of spatiotemporal gait parameter estimation suggesting improved temporal and spatial consistency in comparison to the baseline 3D Pose-only approach.

\begin{table*}[!htbp]
\caption{Correlation coefficient and Mean Absolute Error (MAE) of spatiotemporal gait parameters derived from synthetic markers and pose-estimated heel markers with those obtained from real markers.}
\begin{tabular}{clllllllll}
\hline
\multicolumn{1}{|c|}{}  & \multicolumn{1}{c|}{}  & \multicolumn{2}{c|}{\begin{tabular}[c]{@{}c@{}}Stride \\ time\end{tabular}} & \multicolumn{2}{c|}{\begin{tabular}[c]{@{}c@{}}Stride \\ length\end{tabular}} & \multicolumn{2}{c|}{\begin{tabular}[c]{@{}c@{}}Step \\ time\end{tabular}} & \multicolumn{2}{c|}{\begin{tabular}[c]{@{}c@{}}Step \\ length\end{tabular}} \\ \cline{3-10} 
\multicolumn{1}{|c|}{\multirow{-2}{*}{Method}}  & \multicolumn{1}{c|}{\multirow{-2}{*}{\begin{tabular}[c]{@{}c@{}}2D Pose \\ estimation method\end{tabular}}} & \multicolumn{1}{c|}{corr}    & \multicolumn{1}{c|}{MAE}     & \multicolumn{1}{c|}{corr}     & \multicolumn{1}{c|}{MAE}      & \multicolumn{1}{c|}{corr}   & \multicolumn{1}{c|}{MAE}    & \multicolumn{1}{c|}{corr}    & \multicolumn{1}{c|}{MAE}     \\ \hline
\multicolumn{1}{|c|}{}  & \multicolumn{1}{l|}{MMPose}    & \multicolumn{1}{l|}{0.5457}  & \multicolumn{1}{l|}{0.1061}  & \multicolumn{1}{l|}{0.2957}   & \multicolumn{1}{l|}{0.1628}   & \multicolumn{1}{l|}{0.425}  & \multicolumn{1}{l|}{0.0715} & \multicolumn{1}{l|}{0.3234}  & \multicolumn{1}{l|}{0.1287}  \\ \cline{2-10} 
\multicolumn{1}{|c|}{}  & \multicolumn{1}{l|}{YOLO}      & \multicolumn{1}{l|}{0.5439}  & \multicolumn{1}{l|}{0.0750}  & \multicolumn{1}{l|}{0.4117}   & \multicolumn{1}{l|}{0.1582}   & \multicolumn{1}{l|}{0.3142} & \multicolumn{1}{l|}{0.0425} & \multicolumn{1}{l|}{0.5711}  & \multicolumn{1}{l|}{0.0810}  \\ \cline{2-10} 
\multicolumn{1}{|c|}{\multirow{-3}{*}{3D Pose only}}    & \multicolumn{1}{l|}{OpenPose}  & \multicolumn{1}{l|}{0.5778}  & \multicolumn{1}{l|}{0.1027}  & \multicolumn{1}{l|}{0.5659}   & \multicolumn{1}{l|}{0.1243}   & \multicolumn{1}{l|}{0.5683} & \multicolumn{1}{l|}{0.0700} & \multicolumn{1}{l|}{0.5472}  & \multicolumn{1}{l|}{0.0475}  \\ \hline
\multicolumn{1}{|c|}{}  & \multicolumn{1}{l|}{MMPose}    & \multicolumn{1}{l|}{0.6831}  & \multicolumn{1}{l|}{0.0754}  & \multicolumn{1}{l|}{0.4109}   & \multicolumn{1}{l|}{0.0910}   & \multicolumn{1}{l|}{0.5869} & \multicolumn{1}{l|}{0.0312} & \multicolumn{1}{l|}{0.3691}  & \multicolumn{1}{l|}{0.0554}  \\ \cline{2-10} 
\multicolumn{1}{|c|}{}  & \multicolumn{1}{l|}{YOLO}      & \multicolumn{1}{l|}{0.6642}  & \multicolumn{1}{l|}{0.0487}  & \multicolumn{1}{l|}{0.5717}   & \multicolumn{1}{l|}{0.0811}   & \multicolumn{1}{l|}{0.5807} & \multicolumn{1}{l|}{0.0274} & \multicolumn{1}{l|}{0.7609}  & \multicolumn{1}{l|}{0.0430}  \\ \cline{2-10} 
\multicolumn{1}{|c|}{\multirow{-3}{*}{Proposed method}} & \multicolumn{1}{l|}{OpenPose}  & \multicolumn{1}{l|}{0.7787}  & \multicolumn{1}{l|}{0.0362}  & \multicolumn{1}{l|}{0.7060}   & \multicolumn{1}{l|}{0.0327}   & \multicolumn{1}{l|}{0.8034} & \multicolumn{1}{l|}{0.0195} & \multicolumn{1}{l|}{0.9807}  & \multicolumn{1}{l|}{0.0117}  \\ \hline
\multicolumn{1}{l}{}    &&      &      &       &       &     &     &      &     
\end{tabular}
\vspace{-20pt}
\label{table:spatiotemporal_results1}

\end{table*}

\begin{figure*}[!htbp]
    \centering

    \begin{tabular}{ccc}
        \makebox[0.3\textwidth][c]{\textbf{Knee angle}} 
        & \makebox[0.3\textwidth][c]{\textbf{Pelvis vertical translation}} 
        & \makebox[0.3\textwidth][c]{\textbf{Hip flexion}} \\
        
          
        \begin{subfigure}{0.3\textwidth}          \includegraphics[width=\linewidth]{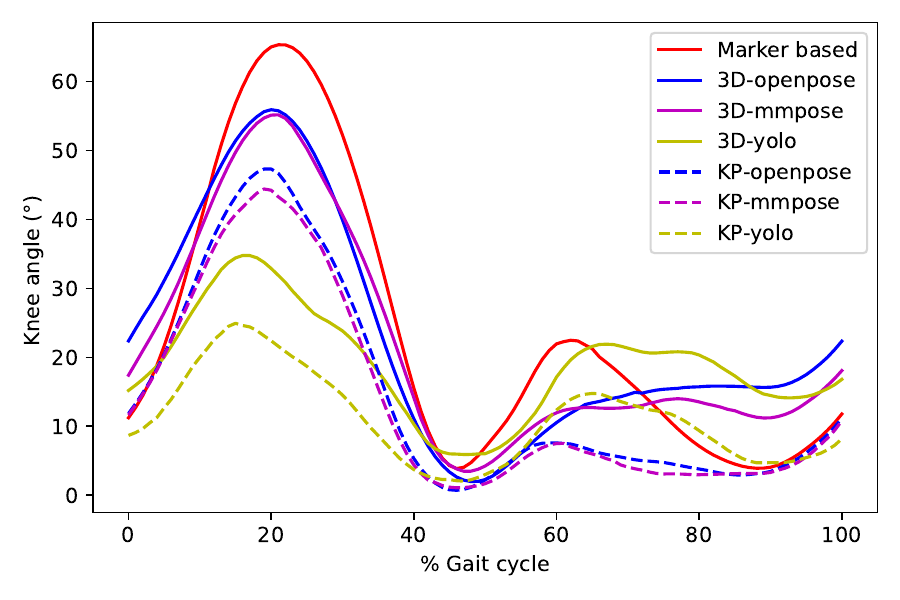}
        \end{subfigure} &
        \begin{subfigure}{0.3\textwidth}
            \includegraphics[width=\linewidth]{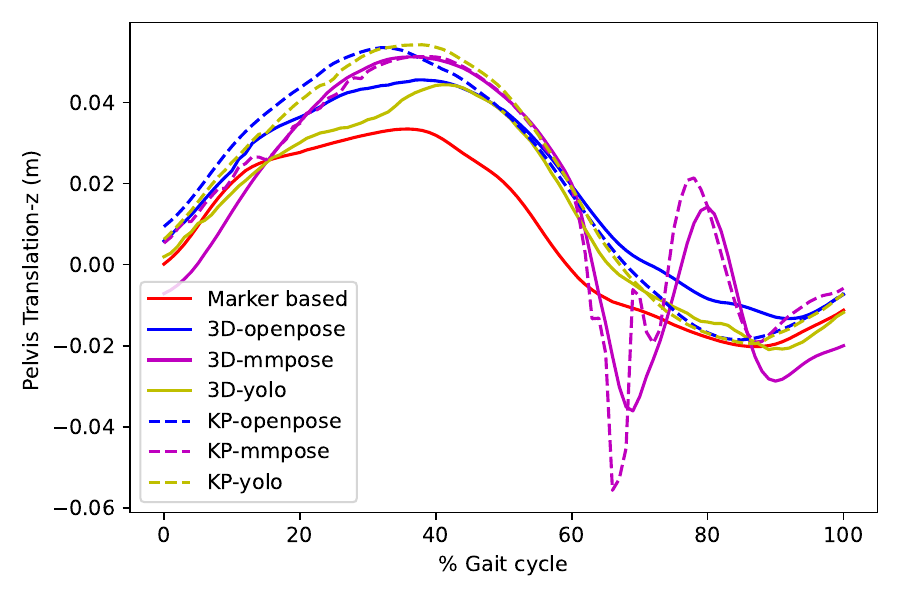}
        \end{subfigure} &
        \begin{subfigure}{0.3\textwidth}
            \includegraphics[width=\linewidth]{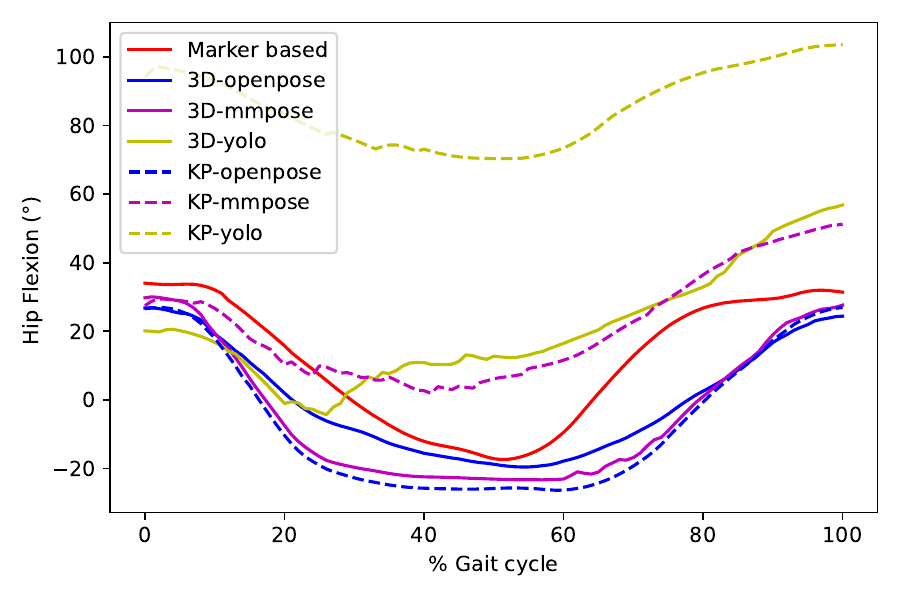}
        \end{subfigure} \\
        
    \end{tabular}
 \vspace{-20pt} 
    \caption{Comparison of kinematic parameters across gait phases.\vspace{-20pt}}
    \label{fig:gait_kinematics}
\end{figure*}

\begin{figure*}[htbp]
    \centering

    \begin{tabular}{ccc}
         \makebox[0.3\textwidth][c]{\textbf{Knee angle}} 
        & \makebox[0.3\textwidth][c]{\textbf{Pelvis vertical translation}} 
        & \makebox[0.3\textwidth][c]{\textbf{Hip flexion}} \\
        
          
        \begin{subfigure}{0.3\textwidth}
            \includegraphics[width=\linewidth]{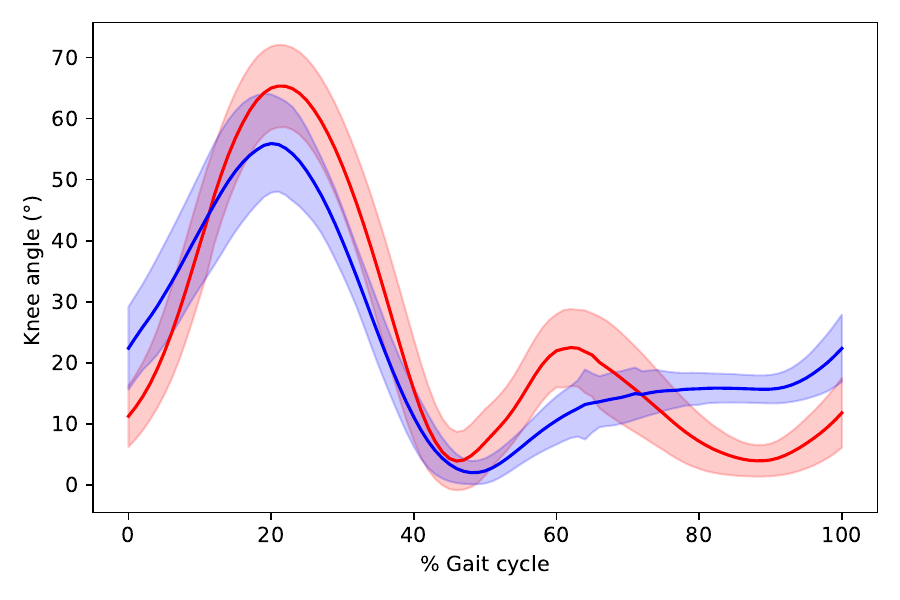}
        \end{subfigure} &
        \begin{subfigure}{0.3\textwidth}
            \includegraphics[width=\linewidth]{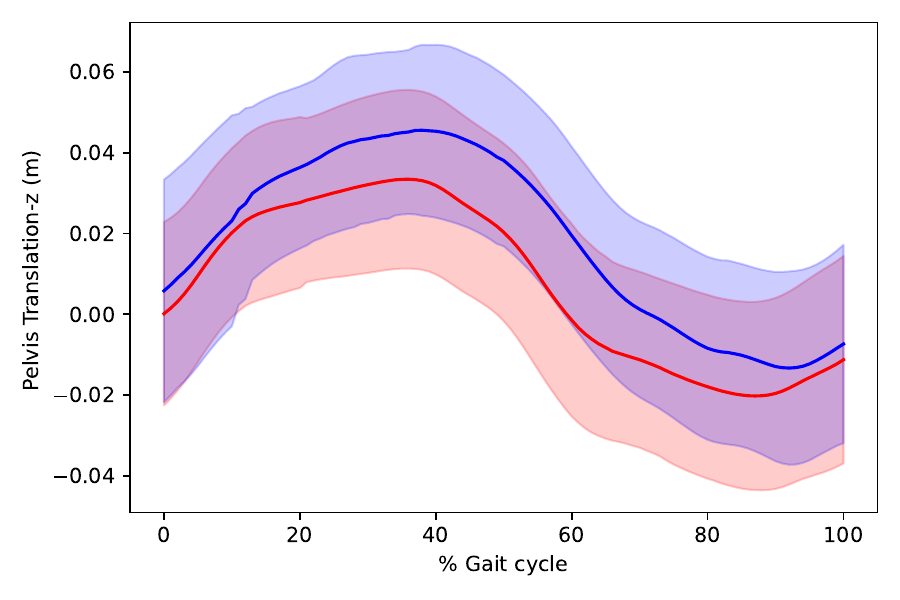}
        \end{subfigure} &
        \begin{subfigure}{0.3\textwidth}
            \includegraphics[width=\linewidth]{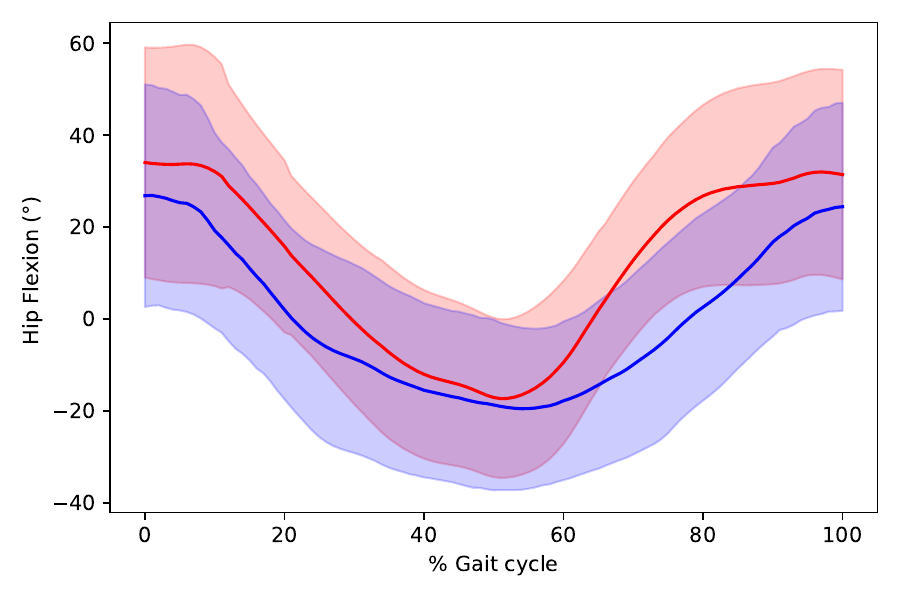}
        \end{subfigure} \\

    \end{tabular}
    \vspace{20pt}
    \begin{tikzpicture}[line width=1pt, x=1cm, y=1cm]

\draw[red!50, line width=6pt, opacity=0.4] (0,0) -- (1,0); 
\draw[red, line width=1.5pt] (0,0) -- (1,0);                
\node[anchor=west] at (1.2,0) {Marker based  Mean $\pm$ SD};

\draw[blue!50, line width=6pt, opacity=0.4] (6.5,0) -- (7.5,0);
\draw[blue, line width=1.5pt] (6.5,0) -- (7.5,0);
\node[anchor=west] at (7.7,0) {Proposed method  Mean $\pm$ SD};

\end{tikzpicture}
 \vspace{-20pt} 
    \caption{Averaged gait cycle waveforms of marker based method and proposed method in its best performing pose estimation method (i.e., OpenPose \cite{cao2019openpose}).}
    \label{fig:mean_and_std_waveform}
\end{figure*}
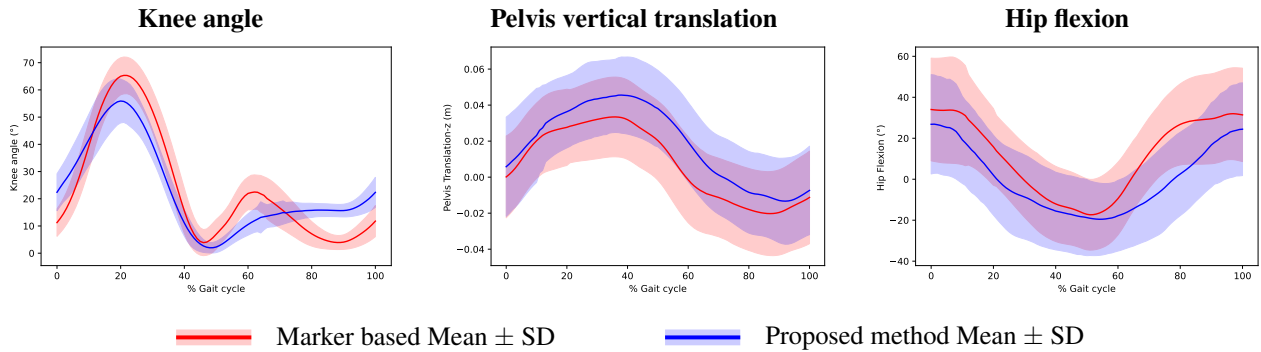

\begin{figure*}[htbp]
    \centering
\vspace*{-20pt}
    \begin{tabular}{ccc}
         \makebox[0.3\textwidth][c]{\textbf{Knee angle}} 
        & \makebox[0.3\textwidth][c]{\textbf{Pelvis vertical translation}} 
        & \makebox[0.3\textwidth][c]{\textbf{Hip flexion}} \\
        
          
        \begin{subfigure}{0.3\textwidth}
            \includegraphics[width=\linewidth]{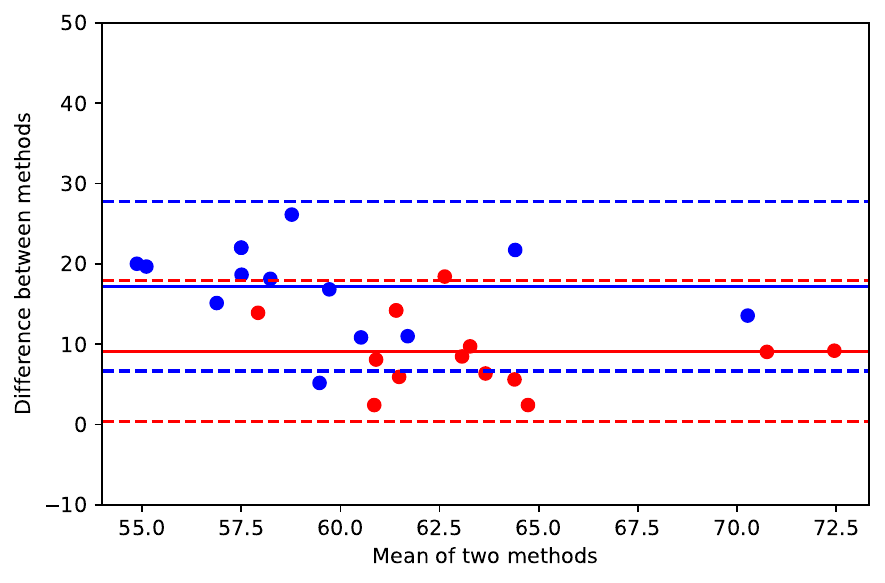}
        \end{subfigure} &
        \begin{subfigure}{0.3\textwidth}
            \includegraphics[width=\linewidth]{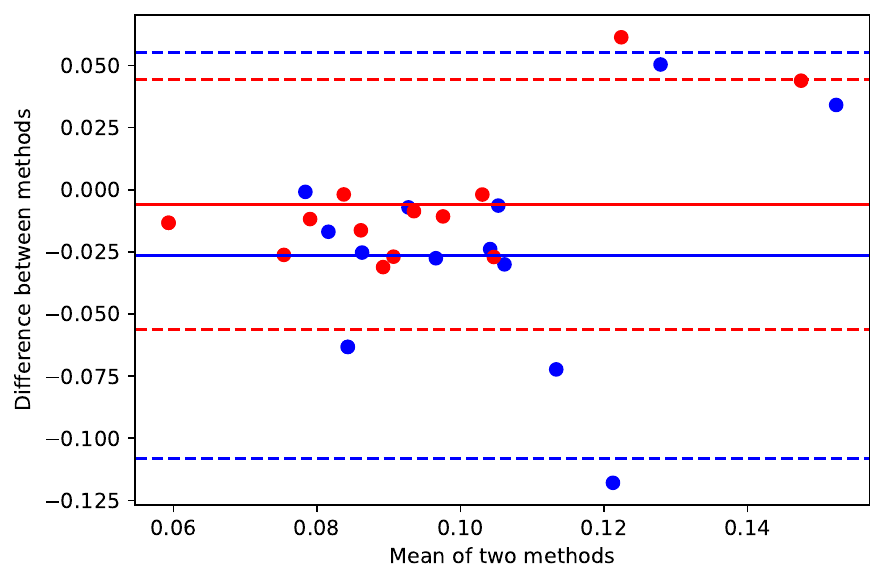}
        \end{subfigure} &
        \begin{subfigure}{0.3\textwidth}
            \includegraphics[width=\linewidth]{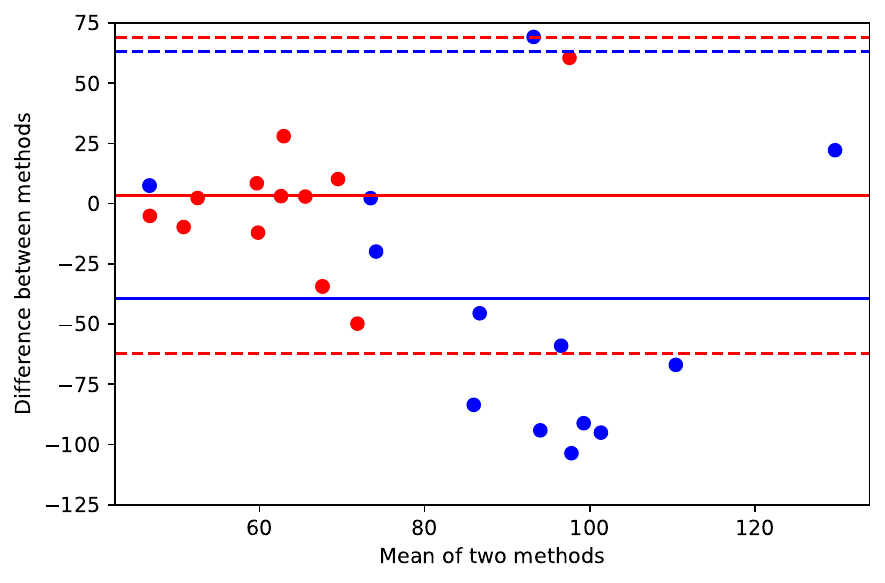}
        \end{subfigure} \\

    \end{tabular}
 \vspace{-10pt} 
    \caption{Bland-Altman plots comparing marker based and markerless joint angle estimations from OpenPose.  \textcolor{red}{Red} denotes the proposed method, and \textcolor{blue} {blue}  denotes the keypoint only method. The solid  lines indicate the mean bias between the two methods, while the dashed lines (--) represent the 95 \% limits of agreement (±1.96 SD).}
    \label{fig:blant-altman}
\end{figure*}

\noindent \textbf{Kinematic gait parameters}: For the kinematic parameter analysis, the joint angles and translations derived from the OpenSim inverse kinematics process were used. 
For the pose-only based method, the 3D keypoints were used as the closest corresponding anatomical markers, and the same inverse kinematics model was applied to estimate joint angles and positions.

Figure \ref{fig:gait_kinematics} presents the mean-time normalized gait cycles for knee angle, pelvis rotation, and hip flexion. From the figure, it can be observed that the joint kinematic patterns obtained using the proposed method exhibit a strong resemblance to those derived from the marker-based reference data. The temporal alignment and curve shape are highly consistent across the gait cycle, indicating that the proposed approach effectively captures both the amplitude and phase characteristics of the underlying motion. Furthermore, the same performance trends observed in the spatiotemporal parameter analysis are reflected here, where the proposed method demonstrates improved agreement and reduced deviation relative to the baseline 3D pose-only approach and the proposed method with OpenPose \cite{cao2019openpose} keypoint extraction shows the best performance.

Following the clinical evaluation process \cite{carvalho2025construct,mansournia2021bland} to evaluate how closely the two measurements align, the mean difference (bias) and 95\% limits of agreement (LoA) were determined following the Bland-Altman analysis. The bias was obtained as the average difference between the experimental marker-based and real marker-based results, while the 95\% LoA were defined as the $ LOA = bias \pm 1.96\ast SD_{diff}$, where $SD_{diff}$ represents the standard deviation of the differences. This analysis was performed on the range-of-motion (ROM) values, defined as the difference between maximum and minimum angles associated with each motion. The results for three joint parameters: knee angle, pelvis rotation, and hip flexion are illustrated in Fig. \ref{fig:blant-altman}. 

Since the OpenPose-based framework demonstrated the highest level of agreement with the marker-based reference in the previous analysis (Fig. \ref{fig:gait_kinematics} and Fig. \ref{fig:mean_and_std_waveform}),  the Bland–Altman plots corresponding to the OpenPose-based method are presented in Fig. \ref{fig:blant-altman}. Each plot illustrates the comparison between the proposed method (red markers) and the keypoint-only method (blue markers) across the three kinematic parameters. The proposed method shows smaller mean biases and narrower 95 \% limits of agreement compared to the keypoint-only approach, indicating improved consistency with the reference data. The tighter clustering of points around the mean line further suggests that extracting markers from the reconstructed 3D structure enhances the accuracy and reliability of kinematic gait parameter estimation.
\vspace{-10pt}

\section{Conclusion}
\vspace{-5pt}
This study presented a biomechanically interpretable framework for gait analysis that bridges the gap between marker based motion capture and modern vision based pose estimation. By reconstructing a detailed 3D human body model and extracting anatomically meaningful virtual markers, the proposed method enables direct integration with biomechanical modeling tools such as OpenSim for kinematic and spatiotemporal gait analysis.

The results demonstrate that the proposed approach achieves strong agreement with reference marker based measurements, as reflected by high correlation coefficients, low mean absolute errors, and reduced bias in Bland-Altman analyses. These findings indicate that deriving biomechanical markers from reconstructed 3D structures substantially improves both the accuracy and consistency of gait parameter estimation compared to  keypoint only approaches. Future work of this analysis will focus on expanding the framework to incorporate a wider range of state-of-the-art pose estimation methods, analysis of other movement types, using more complex 3D shape models and use of this in other clinical end applications.

\section{Compliance with Ethical Standards}
\vspace{-10pt}
This study adhered to the ethical approval EP1819052 
(25/07/2019) for the BioCV dataset and the ethical approval HREC Reference: 2024/86 for the data depicted in Fig. 2 of this paper.

\vspace{-20pt}
\bibliographystyle{IEEEbib}
\bibliography{strings,refs}

\end{document}